\documentstyle[prd,preprint,aps]{revtex}

\begin{document}

\reversemarginpar

\title{The quantum phase problem: steps toward a resolution}
\author{Gilad Gour} 
\address{Racah Institute of Physics, Hebrew University of Jerusalem,\\ 
Givat Ram, Jerusalem~91904, ISRAEL; E-mail:~gour@cc.huji.ac.il}

\maketitle

$\;$\\

Defining the observable ${\bf \phi}$ canonically conjugate to the number
observable ${\bf N}$ has long been an open problem in quantum theory.  The
problem stems from the fact that ${\bf N}$ is bounded from below.  In
a previous work we have shown
how to define the absolute phase observable ${\bf \Phi}\equiv |{\bf\phi}|$ 
by suitably restricting the Hilbert space of $x$ and
$p$ like variables. Here we show that also from the classical point of view,
there is no rigorous definition for the phase even though it's absolute
value is well defined.

\section{Introduction}
Defining the observable ${\bf \phi}$ which represents the phase of a monochromatic
wave has long been an open problem in quantum theory~\cite{Rev}. 
In a previous work~\cite{Gour} the problem has been investigated and solved partly
by the construction of the $absolute$ value ${\bf \Phi}=|{\bf \phi}|$ of the phase
observable which is determined modulus $\pi$ (and not $2\pi$). 
As Moshinsky and Seligman~\cite{MoSe} have shown many years ago, the classical
canonical transformations to action and angle variables for the harmonic oscillator
(and some other systems) turn out to be non-bijective (not one-to-one onto).
Hence, it is not surprising that it is possible to construct only the $absolute$ 
phase observable and not the phase observable itself. The main purpose of this 
paper is to extend the idea and to show that even in the classical point of view 
it is only the absolute value of the phase which is well determined.
Our technique looks
similar to the one given by Newton~\cite{Newton}, however, in his approach 
he doubled the Hilbert space of a simple harmonic oscillator.

Problems in the definition of the quantum phase were first addressed by Fritz 
London~\cite{London} in 1926. One year latter Dirac~\cite{Dirac} 
introduced an operator 
solution which was proved to be incomplete by Susskind and Glogower~\cite{Susskind}
(for history and measurements see~Nieto~\cite{Nieto}).
Since then a series of workers have made many attempts
to resolve the problem (for reviews see~\cite{Rev} and for recent attempts 
see~\cite{Kastrup}).
However, some of the quantum phase theories do not pass the Barnett
and Pegg ``acid-test''~\cite{Barnett}: not all the number states represent states of
random phase. Others suffer from distressing mathematical difficulties. 
A few of the theories are even incomplete, etc. We shall mention that there $is$ a 
solution if one generalize the formal description of measurement to include the 
so-called POM or POVM observables~\cite{Shapiro}. In the previous work it has been 
shown that it is possible to define rigrously the absolute value of the phase
operator without generalizing the formal description of measurement.
 As we shall see in this work, 
the source of all difficulties lies in the fact that the domain of a well defined
phase observable must be restricted to half of the domain which is used in most of 
the quantum phase theories. This is a direct consequence of the fact that  
the number operator ${\bf N}$ is bounded from below.

Until the last years, there were minimal experimental works, compared
to the huge quantity of theoretical effort. We shall mention here the
work of Noh, Foug\`{e}res and Mandel~\cite{Mandel}. By analyzing classical
phase measurement configurations they define the ``operational phase operators''.
There measurements are found to agree very well with the theoretical predictions
of these operators and disagree with the predictions of some of the theories
mentioned in~\cite{Rev}. There are many other experiments~\cite{Rev,Nieto,Exp} 
which have been performed in recent years, however, in this paper we shall focus 
on the theoretical aspect of the quantum phase.  

The phase observable
should be canonically conjugate to the number operator, and thus
represent also the time operator of a simple harmonic oscillator. 
In classical mechanics it is possible to define the canonical conjugate 
to the Hamiltonian of an harmonic oscillator by performing a canonical
transformation on the standard coordinates $q$ and $p$. 
In this paper, a careful analysis of such transformation shows that the 
notion of a phase which is canonically
conjugate to the Hamiltonian is also problematic at the classical
level. Hence, the first step towards a definition of a quantum
phase observable should be the investigation of the problems as it appear
in the classical picture (section II).

The paper is organized as follows. In section II it is shown that the classical
phase is determined up to modulus $\pi$ by the standard coordinates $q$ and $p$
of a linear harmonic oscillator.
In section III we discuss how problems appear in the construction of
a quantum phase and the notion of canonical commutation relation. 
In section IV we construct the absolute quantum phase and examine its properties.
In section V we discuss the quantum phase theories in a finite dimensional
space. Finally, in section VI, we present our summary and conclusions.

\section{Phase in classical theory}

In the Lagrangian formalism, one dimensional classical system can be 
described by one general coordinate $q(t)$ and its time derivative. 
The canonical conjugate to $q(t)$ is defined by
\begin{equation}
p(t)=\frac{\partial L}{\partial\dot{q}(t)}
\label{defc}
\end{equation}
where $L=L(q(t),\dot{q}(t),t)$ is the Lagrangian describing the system.
Hence, $p(t)$ can be written as a function of $q(t)$, $\dot{q}(t)$ and $t$.
In the Hamiltonian formulation, on the other hand, two variables
$q(t)$ and $p(t)$ are said to be canonically conjugate if there exists some function
(Hamiltonian) $H(q,p,t)$ such that the equations of motion are given in the form
\begin{equation}
\dot{q}=\frac{\partial H}{\partial p}\;,\;\;\;\;
\dot{p}=-\frac{\partial H}{\partial q}.
\label{canc}
\end{equation}  
As it will be shown below, in the case of harmonic oscillator there exist
two variables $Q$ (the phase) and $P$ (proportional to the Hamiltonian)
which satisfy Eq.~(\ref{canc}) even though
there is {\it no} Lagrangian $L(Q,\dot{Q},t)$ such that Eq.~(\ref{defc}) 
is satisfied. Hence, $Q$ and $P$ are canonically conjugate according
to the Hamiltonian formalism but not according to the Lagrangian 
formalism. 

The Hamiltonian of a simple harmonic oscillator in one dimension
can be written as
\begin{equation}
H=\frac{1}{2m}(p^{2}+m^{2}\omega^{2}q^{2}),
\end{equation}
where $\omega$ is the frequency of oscillations.
We shall consider here a canonical transformation $q,p\rightarrow Q,P$
such that the Hamiltonian in the new coordinates can be written as 
$H=\omega P$. Thus, $P$ represents the classical analog to the number 
operator. The generating function $F_{1}(q,Q)$ of such a transformation is 
given by~\cite{Goldstein}
\begin{equation}
F_{1}=\frac{m\omega q^{2}}{2}\cot Q.
\end{equation}
However, there are {\it two} transformations derived from $F_{1}$ 
(a generating function may at times be double-valued), namely    
\begin{eqnarray}
\left\{
\begin{array}{ll}
& p=\sqrt{2m\omega P}\cos Q\\
& q=\sqrt{\frac{2P}{m\omega}}\sin Q\\
\end{array}\right.
\;\;\;\;{\rm and}\;\;\;\;
\left\{
\begin{array}{ll}
& p=-\sqrt{2m\omega P}\cos Q\\
& q=-\sqrt{\frac{2P}{m\omega}}\sin Q\;\;.\\
\end{array}\right.
\label{trans}
\end{eqnarray}
The inverse transformations 
\begin{eqnarray}
P=\frac{1}{2m\omega}(p^{2}+m^{2}\omega^{2}q^{2})\;\;\;{\rm and}\;\;\;
Q={\rm cot}^{-1}\left(\frac{p}{m\omega q}\right),
\label{Qdef}
\end{eqnarray}
shows that $q$ and $p$ determine $Q$ modulus $\pi$ and $not$ $2\pi$ because 
a shift in $Q$ by $\pi$ corresponds to going from
one transformation to the other (see Eq.~(\ref{trans})). 
Hence, in order to obtain the classical phase which is determined
modulus $2\pi$, we would have to combine (somewhat artificially) the two 
transformations in Eq.~(\ref{trans}). As we shall see, such a combination
has no quantum analog (and thus the definition of a quantum phase is problematic).
Therefore, we expect the quantum counterpart of $Q$ to be determined by the position 
and momentum operators up to modulus $\pi$.
This explains why it is possible to define only the absolute value of the phase 
observable~\cite{Gour} which   
is restricted to the domain $[0,\pi]$ (assuming the phase 
itself is defined in the interval $(-\pi,\pi]$).

Notice that it is impossible to express $P$ as a function of $Q$ and $\dot{Q}$ 
because $\dot{Q}=\frac{\partial H}{\partial P}=\omega$. Thus, it is impossible
to construct a Lagrangian $L(Q,\dot{Q},t)$ such that 
$P=\frac{\partial L}{\partial\dot{Q}}$. 
This shows that $P$ and $Q$ can be considered as canonically 
conjugated variables only in the Hamiltonian formalism. However,
even then, if $Q$ is 
restricted to the domain $(-\pi,\pi]$ (by taking $Q$ mod $2\pi$) then it is no 
longer the canonical conjugate of $P$; that is: if $Q$ satisfy the Hamilton 
equation $\dot{Q}=\frac{\partial H}{\partial P}=\omega$, then 
$Q=\omega t+\phi_{0}$ can not be restricted! 

How is the fact that the phase is determined
modulus $\pi$ is consistent with the phase which is determined
by the action-angle method? The so called action variable $J$ is defined 
as~\cite{Goldstein}
\begin{equation}
J=\frac{1}{2\pi}\oint pdq
\label{jjarati}
\end{equation}
where the integration is to be carried over a complete period. 
Now, the momentum of a linear harmonic oscillator is given by
\begin{equation}
p=\pm\sqrt{2m\alpha -m^{2}\omega^{2}q^{2}},
\end{equation}
where $\alpha\equiv H$ represents the constant energy of the oscillator.
Hence, substituting the expression for $p$ in Eq.~(\ref{jjarati}) 
yields~\cite{Goldstein}
\begin{equation}
J=\pm\frac{H}{\omega},
\end{equation} 
or solving for the Hamiltonian,
\begin{equation}
H=\pm \omega J.
\label{hamil}
\end{equation}
Hence, $J$ itself cannot represent the classical analog for the number
operator since it is $not$ bounded from below. 

As we can see from 
Eq.~(\ref{hamil}) it is the absolute value of $J$ which is proportional
to the Hamiltonian. The angle variable $w$ is therefore
\begin{equation}
w = \omega\; {\rm sign} (J)\; t + \beta
\end{equation}
which is related to $q$ and $p$ by
\begin{equation}
q=\sqrt{\frac{2|J|}{m\omega}}\sin w\;,\;\;\;\; 
p=\sqrt{2m\omega |J|}\sin w.
\end{equation}
However, in this case the angle variable $w$ is determined up to modulus
$2\pi$ because it depends also on the sign of $J$. This does not conflict
with the fact $Q$ is determined by $q$ and $p$ up to modulus $\pi$ because 
$w$ is the canonical conjugate of $J$ and not of $|J|=H/\omega$.

\section{Problems in the definition of a quantum phase}

In quantum mechanics two observers ${\bf q}$ and ${\bf p}$ are said to be 
canonically conjugate if they satisfy the commutation relation 
$[{\bf q},{\bf p}]=i\hbar$. 
Usually, this definition is equivalent to the definition given in the Lagrange 
formalism
\begin{equation}
{\bf p}=\frac{\partial L({\bf q},\dot{{\bf q}},t)}{\partial\dot{{\bf q}}}.
\end{equation}
As we have seen in the previous section it is impossible to construct a 
Lagrangian for the harmonic oscillator in terms of the phase observable
${\bf\phi}$ (the analog of $Q$) and its time derivative. Hence,
we may use only the Hamiltonian formalism in order to define
the phase observable.

In section II it has also been shown that if $Q$ is restricted to some 
domain, then $Q$ and $P$ do not satisfy the Hamilton equations. This
phenomenon is manifested in the quantum picture by the fact that
$[{\bf N},{\bf\phi}]\neq i$ for a restricted phase operator. However,
if ${\bf N}$ and ${\bf \phi}$ do not satisfy the canonical commutation
relation, then how can the phase observable be defined, and in what sense 
can ${\bf N}$ and ${\bf \phi}$ be regarded as canonically conjugate 
observables? In order to answer these questions we shall first examine
why the phase operator defined many years ago by Dirac cannot be Hermitian.

Dirac proposed to decompose the annihilation operator in the form 
\begin{equation}
{\bf a}=\exp(-i{\bf \phi}){\bf N}^{1/2}.
\label{pdef}
\end{equation}
However, such a decomposition is problematic since the operator
\begin{equation}
{\bf E}\equiv ({\bf N}+1)^{-1/2}{\bf a}=\sum_{n=0}^{\infty}|n\rangle\langle n+1|,
\end{equation}
(which represents $\exp(-i{\bf \phi})$) is not unitary 
(see Susskind and Glogower~\cite{Susskind}).  
Eq.~(\ref{pdef}) seems reasonable in view of
the classical analog 
\begin{equation}
a\equiv\frac{1}{\sqrt{2m\omega}}p
-i\sqrt{\frac{m\omega}{2}}q=\sqrt{P}\exp(-iQ),
\label{cadef}
\end{equation}
where we have used Eq.~(\ref{trans}). However, the above decomposition for $a$ 
is incomplete since $Q$ is determined by $q$ and $p$ modulus $\pi$ and
$Q\in[0,\pi]$ (see section II). That is, the half domain of $Q$ makes it impossible
to cover the domain of $a$. Hence, we have to find a different relation
between $a$ and $Q$. For example, we can use the relation 
\begin{equation}
p=\sqrt{2m\omega P}\cos Q\;\;\Rightarrow\;\;\cos Q=\frac{1}{2\sqrt{P}}(a+a^{*})
\label{cos}
\end{equation}
since $\cos Q$ can have all values in the domain $[-1,1]$ ($Q\in[0,\pi]$).
The quantum analog of Eq.~(\ref{cos}) is given by the operator of 
Carruthers and Nieto~\cite{CandN}:
\begin{equation}
{\bf C}\equiv\frac{1}{2}({\bf E}+{\bf E}^{\dag}).
\end{equation}
However, as we have shown previously~\cite{Gour}, this operator does not
exactly represent
the ``cosine'' of the quantum phase and a {\it small} correction is needed. 
Now, in the domain $Q\in[0,\pi]$, $\sin Q$ is positive and thus we can
use the relation
\begin{equation}
|q|=\sqrt{\frac{2P}{m\omega}}\sin Q\;\;\Rightarrow\;\;
\sin Q=\left|\frac{1}{2i\sqrt{P}}(a^{*}-a)\right|
\label{sin}
\end{equation}
since $\sin Q$ can only have values in the positive domain $[0,1]$.
The quantum analog of Eq.~(\ref{sin}) is given by the absolute value
of the operator of Carruthers and Nieto~\cite{CandN}:
\begin{equation}
|{\bf S}|\equiv\left|\frac{1}{2i}({\bf E}^{\dag}-{\bf E})\right|.
\end{equation} 
Notice that it is $|{\bf S}|$ which is the quantum analog of $\sin Q$.
However, as we have shown previously~\cite{Gour} 
(and as we reiterate in the next section), a small correction to 
$|{\bf S}|$ is needed.

This explains (also classically) why in the theory of Carruthers and Nieto 
the phase observable which is determined from ${\bf C}$ is {\it different} from 
the one determined from ${\bf S}$, and it is equal (up to a small
correction) to the one determined from $|{\bf S}|$. It is clear that 
${\bf C}$ and $|{\bf S}|$ cannot represent the exact cosine and sine of the 
phase observable, respectively, since for example ${\bf C}^{2}+{\bf S}^{2}\neq 1$. 
A new definition is needed. In the next section we shall obtain the correct 
$\sin{\bf\Phi}$ and $\cos{\bf\Phi}$ (${\bf \Phi}\equiv |{\bf\pi}|$). 

Two main problems appear in the construction of a phase operator.
The first one connected with the fact that the phase operator is an angle operator.
Thus it is restricted to a finite interval which is a problem since, for example,
the matrix elements of $[{\bf N},{\bf\phi}]$ 
in the number state basis $|n\rangle$,
\begin{equation}
\langle n|[{\bf N},{\bf \phi}]|n'\rangle=(n-n')\langle n|{\bf \phi}|n'\rangle
\end{equation}
vanish for $n=n'$ because $|\langle n|{\bf\Phi}|n\rangle|$ is bounded. 
This implies that $[{\bf N},{\bf\phi}]\neq i$. Hence, only unrestricted 
operators can satisfy the standard canonical commutation relations.
Recall that classically, the pair $Q$ mod $\pi$ and $P$ do not satisfy 
the Hamilton equations (even though $Q$ itself and $P$ do). 

The solution to this problem is simple since 
only a slight change in the commutation relations is needed.
For example, in the case of a plane rotator, Judge and Lewis~\cite{Judge} 
showed many years ago that the angular momentum ${\bf J}_{z}$ component
and its associated angle ${\bf\Theta}$ satisfy the commutation relation
\begin{equation}
[{\bf J}_{z},{\bf \Theta}]=i\hbar\left(1-2\pi\delta({\bf \Theta}-\pi)\right),\;\;\;\;
-\pi<\Theta\leq\pi,
\label{npcr}
\end{equation}
where ${\bf\Theta}$ can be expressed as a $2\pi$-periodic function of an 
unrestricted angle operator~\cite{Rev}.

However, Eq.~(\ref{npcr}) can not be used for the case of number and phase 
operators. In addition to the fact that ${\bf \phi}$ is restricted, 
a second problem arises because the number
operator ${\bf N}$ is bounded from below (not so ${\bf J}_{z}$). Therefore, 
a commutation relation like~(\ref{npcr}) for ${\bf N}$ and $\phi$ 
does not hold~\cite{Gour}. For example, by~(\ref{npcr}) 
the matrix elements of $[{\bf N},{\bf \phi}]$ taken in the phase basis
$|\phi\rangle$ (assuming $\phi_{0}<\phi\leq\phi_{0}+\Delta$), would be
\begin{equation}
\langle\phi|[{\bf N},{\bf \Phi}]|\phi'\rangle
=(\phi'-\phi)\langle \phi|{\bf N}|\phi'\rangle=i\delta(\phi-\phi'),
\end{equation}
which imply that
\begin{equation}
\langle \phi|{\bf N}|\phi'\rangle=-i\frac{\delta(\phi-\phi')}{\phi-\phi'}
=i\frac{d}{d\phi}\delta(\phi-\phi').
\end{equation}
Defining a state $|\psi\rangle=\int_{\phi_{0}}^{\phi_{0}+\Delta}d\phi\;
\psi(\phi)|\phi\rangle$
in the basis of the phase states ($\psi(\phi)$ is a complex function of $\phi$), 
we find 
\begin{equation}
\langle\psi|{\bf N}|\psi\rangle=\int_{\phi_{0}}^{\phi_{0}+\Delta}d\phi
\int_{\phi_{0}}^{\phi_{0}+\Delta}d\phi'\;
\psi^{*}(\phi)\psi(\phi')\langle \phi|{\bf N}|\phi'\rangle
=\int_{\phi_{0}}^{\phi_{0}+\Delta}d\phi\;\psi^{*}(\phi)\left(i\frac{d}{d\phi}\right)\psi(\phi).
\label{aven}
\end{equation}
Thus, for the state $\psi(\phi)=\frac{1}{\Delta}\exp(-in\phi)$, desired result
$\langle\psi|{\bf N}|\psi\rangle=n$, but for 
$\psi(\phi)=\frac{1}{\Delta}\exp(+in\phi)$ Eq.~(\ref{aven}) implies a negative average
$\langle\psi|{\bf N}|\psi\rangle=-n$. Thus, Eq.~(\ref{npcr}) cannot hold for
an operator bounded from below.

As shown in section II, the action variable $J$ of a classical
harmonic oscillator is $not$ bounded from below. Thus, its canonical conjugate angle 
variable $w$ is well defined. This explains why the problem of defining the phase
observable of an harmonic oscillator appears only at the quantum level. 
Furthermore, in section II we showed, that the domain of the classical phase $Q$ is 
determined modulus $\pi$ and $not$ $2\pi$. In the next section we shall see
that in the quantum case this follows from the fact that ${\bf N}$ is bounded 
from below.

\section{The quantum phase}

Starting from the position and momentum variables, $q$ and $p$, and then
applying the canonical transformation~(\ref{trans}) enables one to define
the classical phase of the harmonic oscillator (see Eq.~(\ref{Qdef})). In the
quantum picture we start also with the two canonically conjugate operators
${\bf x}$ and ${\bf p}$. However, here these two observables reside in a 
one dimensional Hilbert space ${\cal H}$ of a {\it free} particle. 
Then, in order to obtain the spectrum of a number operator for ${\bf p}$,
one must require for periodicity in $x$ space and symmetry under reflection in $p$
space.

\subsection{The periodic subspace ${\cal H}_{L}$}
    
The periodic subspace ${\cal H}_{L}$ is defined by the requirement that for any
$|\psi\rangle\in {\cal H}_{L}$ and for any position eigenstate 
$|x\rangle\in {\cal H}$, $\langle x-L/2|\psi\rangle=\langle x+L/2|\psi\rangle$.
Since a position shift of $L$ does not alter the states in ${\cal H}_{L}$
we can restrict $x$ to the range $(-\L/2,L/2]$. Furthermore, the momentum 
eigenstates $|p_{n}\rangle$ in ${\cal H}_{L}$ have eigenvalues
\begin{equation} 
p_{n}\equiv \frac{2\pi\hbar n}{L}\;\;\;n=0,\pm 1,\pm 2,...
\end{equation}
Thus the position and momentum observables in ${\cal H}_{L}$
can be written as
\begin{equation}
{\bf x}_{L}=\int_{-L/2}^{L/2}x\;|x\rangle\langle x|\;dx\;\;\;\;\;\;
{\bf p}_{L}=\sum_{n=-\infty}^{\infty}p_{n}\;|p_{n}\rangle\langle p_{n}|,
\end{equation}
where the subscript $L$ indicates that these observables describe 
a particle in a box of size $L$.
Now, since the position and momentum eigenstates  
$|x\rangle,|p\rangle\in{\cal H}$ satisfy the relation
\begin{equation}
\langle x|p\rangle=(2\pi)^{-1/2}\exp\left(-ixp/\hbar\right),
\label{xp1}
\end{equation}
then $|x\rangle,|p_{n}\rangle\in{\cal H}_{L}$ satisfy the relation
\begin{equation}
\langle x|p_{n}\rangle=L^{-1/2}\exp\left(-ixp_{n}/\hbar\right),
\label{ccon}
\end{equation}
where the normalization is in accordance with the domain of $x\in(-L/2,L/2]$.
However, equation Eq.~(\ref{ccon}) implies that the commutation relation
between ${\bf x}_{L}$ and ${\bf p}_{L}$ is no longer $i\hbar$. This is not 
surprising because the analogous thing happens also in the classical picture. 
As we have shown in section~II, if $Q$ and $P$ (see there definitions in 
Eq.~(\ref{Qdef})) satisfy the Hamilton equations then $Q$ must be
unrestricted. This means that a restricted $Q$ (i.e. $Q$ mod $2\pi$) is not
canonically conjugate to $P$. Thus, a restricted position operator ${\bf x}_{L}$
and the momentum operator ${\bf p}_{L}$ satisfy a non-standard commutation
relation:
\begin{eqnarray}
[{\bf p}_{L},{\bf x}_{L}] & = & 
\sum_{n,n'=-\infty}^{\infty}|p_{n}\rangle\langle p_{n}|
({\bf p}_{L}{\bf x}_{L}-{\bf x}_{L}{\bf p}_{L})
|p_{n'}\rangle\langle p_{n'}|\nonumber\\
& = & \frac{1}{L}\sum_{n,n'=-\infty}^{\infty}(p_{n}-p_{n'})
|p_{n}\rangle\langle p_{n'}|\int_{-\frac{L}{2}}^{\frac{L}{2}}xe^{\frac{i}
{\hbar}(p_{n}-p_{n'})x}dx\nonumber\\
& = & i\hbar\sum_{n,n'=-\infty}^{\infty}
\left[\delta_{n,n'}-e^{-\frac{iL}{2\hbar}(p_{n}-p_{n'})}\right]
|p_{n}\rangle\langle p_{n'}|\nonumber\\
& = & i\hbar\left(1-L\left|x=L/2\right\rangle
\left\langle x=L/2\right|\right)
=i\hbar\left(1-L\delta\left({\bf x}-\frac{L}{2}\right)\right).
\end{eqnarray}
This commutation relation also implies a novel uncertainty principle.
That is, for a general state 
\begin{equation}
|\psi\rangle=\sum_{n=-\infty}^{\infty}\psi_{n}|p_{n}\rangle
\end{equation}
in ${\cal H}_{L}$ we find (using the fact that 
$\langle x=L/2|p_{n}\rangle =L^{-1/2}(-1)^{n}$)
\begin{eqnarray}
(\Delta {\bf x}_{L})_{\psi}(\Delta {\bf p}_{L})_{\psi} & \geq &
\frac{1}{2} \left | \langle[{\bf p}_{L},{\bf x}_{L}]\rangle _{\psi}\right|
=\frac{\hbar}{2}\left|1-L\langle\psi|\delta\left({\bf x}-\frac{L}{2}\right)|\psi
\rangle\right|\nonumber\\
& = & \frac{\hbar}{2}\left|1-\left|\sum_{n=-\infty}^{\infty}(-1)^{n}\psi_{n}
\right|^{2}\right|,
\label{uncer}
\end{eqnarray}
where $(\Delta {\bf x}_{L})_{\psi}\equiv\sqrt{\langle\psi|{\bf x}^{2}|\psi\rangle
-(\langle\psi|{\bf x}|\psi\rangle)^{2}}$. Hence, Eq.~(\ref{uncer}) shows clearly
why ${\bf x}_{L}$ is not canonically conjugate to ${\bf p}_{L}$ in the 
{\it strong} standard way. However, in this paper we shall call ${\bf x}_{L}$
and ${\bf p}_{L}$ canonically conjugate observables because of their physical
interpretation: they describe the position and momentum of a particle in a box
of size $L$. But note that it is possible to define a dense {\it subspace} 
${\cal C}\subset{\cal H}_{L}$
\begin{equation}
{\cal C}:\;|\psi\rangle=\sum_{n=-\infty}^{\infty}\psi _{n}|p_{n}\rangle\;,\;\;
{\rm such\;that}\;\; \sum_{n=-\infty}^{\infty}\psi _{n}(-1)^{n}=0,
\end{equation}
in which $[{\bf p}_{L},{\bf x}_{L}]=i$. That is, in the subspace ${\cal C}$,
${\bf x}_{L}$ and ${\bf p}_{L}$ satisfy the strong form of the canonical
commutation relation.

The dimensionless operator ${\bf \Theta}\equiv\frac{2\pi}{L}{\bf x}_{L}$
and the operator ${\bf J}_{z}\equiv\frac{L}{2\pi}{\bf p}_{L}$ can be interpreted
as the angle and angular momentum observables of the plan rotator.
Thus the canonically conjugate to the angular momentum of a plan rotator
is given by the matrix elements of ${\bf \Theta}$ in the angular momentum state 
basis $|j_{n}\rangle\equiv|p_{n}\rangle$
\begin{equation}
\langle j_{n}|{\bf \Theta}|j_{n'}\rangle=\frac{2\pi}{L}\int_{-\frac{L}{2}}^{\frac{L}{2}}
x\langle p_{n}|x\rangle\langle x|p_{n'}\rangle dx
=\frac{i(-1)^{n-n'}}{n-n'}(1-\delta_{nn'}),
\label{matp}
\end{equation}
where we have used Eq.~(\ref{ccon}). The same matrix elements have been obtained 
by Galindo~\cite{Galindo} for the phase observable, but as we see here, these matrix
elements describe the canonical conjugate to the angular momentum observable,
and {\it not} to the number operator.

In order to obtain a positive number-momentum operator we have to add
a further restriction on ${\cal H}_{L}$. Recall that the position $x$-space
has been divided into infinite identical boxes each of size $L$, and then,
since a position shift of $L$ does not alter the states in ${\cal H}_{L}$
it was possible to restrict $x$ to the domain $(-\L/2,L/2]$. In exactly the 
same way we can divide the momentum $p$-space into two identical half spaces 
and then restrict $p$ to the domain $p\geq 0$.

\subsection{The subspace ${\cal H}^{+}_{L}$}

The subspace ${\cal H}^{+}_{L}\subset{\cal H}_{L}$ is defined as follows:
if $|\psi\rangle\in{\cal H}^{+}_{L}$ then 
$\langle p_{n}|\psi\rangle=\langle p_{-n}|\psi\rangle$ for any 
$|p_{n}\rangle\in{\cal H}_{L}$. Hence the momentum
$p$-space in ${\cal H}^{+}_{L}$ is divided into two identical half spaces, namely,
$p\geq 0$ and $p\leq 0$. However, notice that $|p_{n}\rangle$ itself do not belong
to ${\cal H}^{+}_{L}$ and a different definition for a momentum eigenstate is needed.
The new momentum eigenstates are defined by
\begin{equation}
|p_{n}\rangle ^{+}\equiv\frac{1}{2}(|p_{n}\rangle +|p_{-n}\rangle ) 
\label{pplus}
\end{equation}
since ${}^{\;+}\langle p_{n}|\psi\rangle=\langle p_{n}|\psi\rangle$ for 
any $|\psi\rangle\in{\cal H}^{+}_{L}$. Note that any $|\psi\rangle$ in 
${\cal H}^{+}_{L}$ can be written as a superposition of $|p_{n}\rangle ^{+}$'s
states, that is ${\cal H}^{+}_{L}={\rm span}\{|p_{n}\rangle ^{+}\}$.
We shall now examine the $x$-space of ${\cal H}^{+}_{L}$.

For any $|\psi\rangle\in{\cal H}^{+}_{L}$ and for any $|x\rangle\in{\cal H}_{L}$,  
$\langle x|\psi\rangle=\langle -x|\psi\rangle$. This follows because 
$|\psi\rangle$ can be written as 
$|\psi\rangle=\int_{-L/2}^{L/2}dx\;\psi(x)|x\rangle$ since $|\psi\rangle$
also belongs to ${\cal H}_{L}$. The requirement 
$\langle p_{n}|\psi\rangle=\langle p_{-n}|\psi\rangle$ implies that
\begin{equation}
\int_{-L/2}^{L/2}dx\;\psi(x)\exp(ip_{n}x/\hbar)
=\int_{-L/2}^{L/2}dx\;\psi(x)\exp(-ip_{n}x/\hbar)
\end{equation}
for all $n$ and, therefore, $\psi(x)=\psi(-x)$.
Thus also the eigenstates of ${\bf x}_{L}$ itself do not belong to 
${\cal H}^{+}_{L}$. 
The new position eigenstates are defined analogously to 
Eq.~(\ref{pplus})
\begin{equation}
|x\rangle ^{+}\equiv\frac{1}{2}(|x\rangle +|-x\rangle) 
\end{equation}
since ${}^{\;+}\langle x|\psi\rangle=\langle x|\psi\rangle$ for 
any $|\psi\rangle\in{\cal H}^{+}_{L}$.

As we have seen above the Hilbert space ${\cal H}^{+}_{L}$ is symmetric
both in the position and momentum space; it does not distinguish
between $x$ and $-x$ and between $p$ and $-p$. Hence, the orthonormality
conditions of the bases $\{|x\rangle ^{+}\}$ and $\{|p_{n}\rangle ^{+}\}$
are given by
\begin{equation}
{}^{\;+}\langle x'|x\rangle ^{+}=\frac{1}{2}\left(\delta(x-x^{\prime})
+\delta(x+x^{\prime})\right)\;\;\;{\rm and}\;\;\; ^{+}\langle p_{n}|p_{n'}\rangle^{+}
=\frac{1}{2}\left(\delta_{n,n'}+\delta_{n,-n'})\right).
\end{equation}
Furthermore, in a symmetric space the projection of $|p_{n}\rangle^{+}$ on 
$|x\rangle^{+}$ is no longer like Eq.~(\ref{ccon}) since it should be an 
even function of both $x$ and $p$. That is,
\begin{equation}
^{+}\langle x|p_{n}\rangle^{+}=\frac{1}{4}\left(\langle x|p_{n}\rangle
+\langle -x|p_{n}\rangle+\langle x|-p_{n}\rangle+\langle -x|-p_{n}\rangle\right)
=\sqrt{1/L}\cos\left(xp_{n}/\hbar\right),
\end{equation}
where we have used Eq.~(\ref{ccon}).

The next step is to restrict the Hilbert space ${\cal H}^{+}_{L}$ to
the domain of $p\geq 0$ and $x\geq 0$. However, since ${\cal H}^{+}_{L}$
is already symmetric we just have to change the normalization conditions
to be in accordance with the new domain of $x$ and $p$. That is,
\begin{equation}
|x\rangle^{+}\longrightarrow\sqrt{2}|x\rangle^{+}\;\;\;{\rm and}\;\;\;
|p_{n}\rangle^{+}\longrightarrow\sqrt{2}|p_{n}\rangle^{+}
\end{equation}
for $x$ greater then zero and $n\geq 1$. For $|x=0\rangle^{+}$ and
$|p_{n=0}\rangle^{+}$ no change is needed. With these changes, 
the position and momentum observables in the half bounded space $0\leq x\leq L/2$
are given by 
\begin{equation}
{\bf x}^{+}_{L}=\int_{0}^{L/2}x |x\rangle^{+\;+}\langle x|dx\;\;\;
{\rm and}\;\;\;
{\bf p}^{+}_{L}=\sum_{n=0}^{\infty}p_{n}\;|p_{n}\rangle^{+\;+}\langle p_{n}|,
\end{equation}
where the orthonormality conditions in half space are given by
\begin{eqnarray}
{}^{\;+}\langle x|x^{\prime}\rangle^{+} & = & \delta(x-x^{\prime})
\;\;\;{\rm for}\;\;\;x,x^{\prime}>0\;\;\;{\rm and}\;\;\;
^{\;+}\langle x|x^{\prime}=0\rangle^{+}=2\delta(x)\nonumber\\
^{\;+}\langle p_{n}|p_{n'}\rangle^{+} & = & \delta_{n,n'}
\;\;\;{\rm for}\;\;\;n,n'\geq 0.
\end{eqnarray}
The projection of $|p_{n}\rangle ^{+}$ on $|x\rangle^{+}$ in the half space
is given by:
\begin{equation}
^{+}\langle x|p_{n\geq 1}\rangle^{+}=\sqrt{4/L}\cos\left(xp_{n}/\hbar\right)\;\;\;
{\rm and}\;\;\;^{+}\langle x|p_{n=0}\rangle^{+}=\sqrt{2/L}.
\label{pofxonp}
\end{equation}

\subsection{The number and absolute value of phase observables}

The dimensionless number and absolute value of 
phase observables are defined by
\begin{equation}
{\bf N}\equiv\frac{L}{2\pi\hbar}{\bf p}^{+}_{L}=
\sum_{n=0}^{\infty}n|n\rangle\langle n|\;\;\;
{\rm and}\;\;\;{\bf\Phi}\equiv\frac{2\pi}{L}{\bf x}^{+}_{L}
=\int_{0}^{\pi}\phi|\phi\rangle\langle\phi|d\phi,
\label{defp}
\end{equation}
where $|n\rangle\equiv|p_{n}\rangle^{+}$ and 
$|\phi\rangle\equiv\sqrt{\frac{L}{2\pi}}|x=\frac{L}{2\pi}\phi\rangle^{+}$.
In a dimensionless form, Eq.~(\ref{pofxonp}) can be written as
\begin{equation}
|\phi\rangle=\frac{1}{\sqrt{\pi}}|n=0\rangle+\sqrt{\frac{2}{\pi}}
\sum_{n=1}^{\infty}\cos(n\phi)|n\rangle,
\label{phin}
\end{equation}
which is consistent with the normalization 
conditions $\langle\phi|\phi'\rangle=\delta(\phi-\phi')$ for $0<\phi,\phi'\leq\pi$
and $\langle\phi|\phi=0\rangle=2\delta(\phi)$. Furthermore, Eq.~(\ref{phin})
together with the definition of the phase observable~(\ref{defp}) implies
that the matrix elements of ${\bf\Phi}$ in the $|n\rangle$ basis are:
\begin{equation}
\langle n|{\bf\Phi}|n'\rangle=\langle n'|{\bf\Phi}|n\rangle=
\left\{
\begin{array}{ll}
&\pi/2\;\;\;\;{\rm for}\;\;n'=n\\
& -2\sqrt{2}\pi^{-1}n^{-2}
\;\;\;\;{\rm for}\;\;n'=0,\;\; {\rm odd}\;\;n\\
& -2\pi^{-1}
\left((n+n')^{-2}+(n-n')^{-2}\right)\;\;\;{\rm for}
\;\;\;n,n'>0,\;\;{\rm odd}\;\; n+n'\\
&0\;\;\;{\rm otherwise}.
\end{array}
\label{gg}
\right.
\end{equation}

We shall now examine the absolute value of the phase observable 
${\bf\Phi}\equiv |\phi|$ defined by Eq.~(\ref{defp}) and 
Eq.~(\ref{gg}). First, the range of the ${\bf\Phi}$ is from $0$ to $\pi$.
Although the unrestricted phase operator can take values from $-\infty$
to $+\infty$, the values which correspond to actual states will be in this $\pi$
range. This domain of the phase observable is in a complete agreement with
the classical picture, especially with Eq.~(\ref{Qdef}). The fact that the number
operator cannot have negative eigenvalues implies that the phase is also bounded to
half range (i.e. it is possible to define only the absolute value of the phase 
observable). 

The second characteristic of ${\bf\Phi}=|\phi|$ is that $\phi$ itself 
is complementary to the number operator. That is, all the number states, 
including the vacuum, are states of 
random phase $\phi$. This is reasonable since a number state should not have a preferred
phase. This idea is known as the Barnett and Pegg ``acid-test''~\cite{Barnett}
for quantum phase theories. In our theory, the distribution of the phase in a 
number state $|n\rangle$ is given by
\begin{eqnarray}
P_{n=0}(\phi) & \equiv & |\langle n=0|\phi\rangle|^{2}=\frac{1}{\pi}\nonumber\\
P_{n\geq 1}(\phi) & \equiv & |\langle n|\phi\rangle|^{2}=\frac{2}{\pi}\cos^{2}(n\phi)= 
\frac{1}{\pi}(1+\cos(2n\phi)),
\label{nln}
\end{eqnarray}
where we have used Eq.~(\ref{phin}).
Not surprisingly this distribution is not uniform since it is the 
distribution of the absolute value of the phase, and not of the phase itself.
In the case of the plane rotator for example, the state 
$|\theta>_{+}\equiv 1/\sqrt{2}(|\theta>+|-\theta>)$ is an eigenstate of the 
absolute value of the angle operator (I owe this point to M. J. W. Hall), 
and it gives a similar distribution
as in Eq.~(\ref{nln}). 
Thus, Eq.~(\ref{nln}) yields
a non-uniform phase distribution. However, in the classical limit 
$n\rightarrow\infty$, the average of $\Phi^{m}$ ($m=0,1,2...$) in a number 
state is given by:
\begin{equation}
\langle n| {\bf\Phi}^{m}|n\rangle\equiv\langle\Phi^{m}\rangle_{n}
=\frac{1}{\pi}\int_{0}^{\pi}d\Phi(1+\cos(2n\Phi))\Phi^{m}
\;\rightarrow\;\frac{1}{\pi}\int_{0}^{\pi}d\Phi\Phi^{m}
\label{avnu}
\end{equation}
which is identical to the average
of $\Phi^{m}$ in a classical uniform phase distribution $P(\Phi)=1/\pi$.

The third characteristic of ${\bf\Phi}$~(\ref{defp}) is that it 
represents the absolute value of the non-Hermitian phase operator defined
long time ago by Dirac~\cite{Dirac}. The exponent of this operator can be 
represented by the Susskind and Glogower~\cite{Susskind} non-unitary operators 
\begin{equation}
{\bf E}\equiv ({\bf N}+1)^{-1/2}{\bf a}=\sum_{n=0}^{\infty}|n\rangle\langle n+1|
\;,\;\;\;\;
{\bf E}^{\dag}\equiv {\bf a}^{\dag}({\bf N}+1)^{-1/2}
=\sum_{n=0}^{\infty}|n+1\rangle\langle n|.
\end{equation} 
In order to compare these operators to our approach we shall use 
the definition of Carruthers and Nieto~\cite{CandN} for the Hermitian sine
${\bf S}\equiv\frac{1}{2i}({\bf E}-{\bf E}^{\dag})$ and cosine  
${\bf C}\equiv\frac{1}{2}({\bf E}+{\bf E}^{\dag})$ operators. Note that 
${\bf S}$ and ${\bf C}$ cannot represent the exact sine and cosine of the
phase observables since, for example, $[{\bf S},{\bf C}]\neq 0$. However, 
in states of large average number occupations, 
$\langle {\bf N}\rangle\gg 1$, ${\bf S}$ and ${\bf C}$ can be treated  
approximately as the sine and cosine of the phase. Thus, our theory should produce
(small) corrections to ${\bf S}$ and ${\bf C}$. Since we have the 
absolute value of the phase in the range $[0,\pi]$, the sine of ${\bf\Phi}$ 
is always positive, and 
therefore should be compared with the absolute value of ${\bf S}$. This has been 
explained using the classical picture in section III.

The cosine of the phase operator is defined by:
\begin{equation}
\cos{\bf\Phi} = \int_{0}^{\pi}\cos\phi\;|\phi\rangle\langle\phi|\;d\phi.
\end{equation}
Taking its matrix element in the $|n\rangle$ basis gives
\begin{equation}
\langle n|\cos{\bf\Phi}|n'\rangle =
\int_{0}^{\pi}\cos\phi\;\langle n|\phi\rangle\;\langle\phi|n'\rangle\;d\phi
=\frac{1}{2}\left(\delta_{n,n'+1}+\delta_{n,n'-1}\right)
+\frac{1}{2}(\sqrt{2}-1)\delta_{1,n+n'}\;,
\label{coss}
\end{equation}
where we have used Eq.~(\ref{phin}) for the value of $\langle n|\phi\rangle$.
Now, since the first term with the parenthesis in Eq.~(\ref{coss}) is exactly the 
matrix element $\langle n|{\bf C}|n'\rangle$, the cosine of ${\bf \Phi}$ can
be written in the form
\begin{equation}
\cos{\bf\Phi}={\bf C}+\frac{1}{2}(\sqrt{2}-1)(|0\rangle\langle 1|
+|1\rangle\langle 0|).
\label{1c}
\end{equation}
where the projectors involving the number eigenstates 
$|0\rangle$ and $|1\rangle$ can be neglected 
for states with $\langle {\bf N}\rangle\gg 1$.

In order to compare $|{\bf S}|$ with the sine of ${\bf \Phi}$,
it is enough to find the relation between the square of these operators.
In this way we avoid the need to calculate the absolute value of ${\bf S}$.  
Using the same technique as for $\cos{\bf\Phi}$ we find that
\begin{equation}
\sin^{2}{\bf\Phi} = \int_{0}^{\pi}\sin^{2}\phi\;|\phi\rangle\langle\phi|\;d\phi
={\bf S}^{2}+\frac{1}{4}(1-\sqrt{2})(|0\rangle\langle 2|+|2\rangle\langle 0|)
+\frac{1}{4}(|0\rangle\langle 0|-|1\rangle\langle 1|), 
\label{2s}
\end{equation}
where we have used Eq.~(\ref{phin}). In our theory it is obvious from its definition
that $\sin{\bf\Phi}$ and $\cos{\bf\Phi}$ commute and satisfy the trigonometric
relation $\sin^{2}{\bf\Phi}+\cos^{2}{\bf\Phi}=1$.

\subsection{The classical limit}

We shall show that our quantum phase is in a complete agreement
with the classical limit. The coherent states which correspond to the classical 
picture are the eigenstates of the annihilation operator ${\bf a}$. A
coherent state can be written as
\begin{equation}
|\gamma\rangle=\exp(-\frac{1}{2}|\gamma|^{2})\sum_{n=0}^{\infty}\frac{\gamma^{n}}
{\sqrt{n!}}|n\rangle, 
\end{equation}
where $\gamma=\sqrt{N}e^{i\theta}$ is the eigenvalue of the annihilation
operator ${\bf a}$. Thus, the average of ${\bf \Phi}$ in a coherent state
can be written as 
\begin{eqnarray}
\langle\gamma|{\bf\Phi}|\gamma\rangle & = & 
\exp(-|\gamma|^{2})\sum_{n,n'=0}^{\infty}\frac{\gamma ^{n'} (\gamma ^{*}) ^{n}}
{\sqrt{n!n'!}}\langle n'|{\bf\Phi}|n\rangle\nonumber\\
& = & \exp(-|\gamma|^{2})\sum_{n=0}^{\infty}\frac{|\gamma| ^{2n}}
{n!}\langle n|{\bf\Phi}|n\rangle
+\exp(-|\gamma|^{2})\sum_{n\neq n'}\frac{\gamma ^{n'} (\gamma ^{*}) ^{n}}
{\sqrt{n!n'!}}\langle n'|{\bf\Phi}|n\rangle\nonumber\\
& = & \frac{\pi}{2}+\exp(-|\gamma|^{2})\sum_{n=0}^{\infty}\sum_{s=1}^{\infty}
\left[\frac{|\gamma|^{2n}\gamma ^{s}}{\sqrt{n!(n+s)!}}+
\frac{|\gamma|^{2n}\gamma ^{-s}}{\sqrt{n!(n-s)!}}\right]
\langle n+s|{\bf\Phi}|n\rangle
\label{abba}
\end{eqnarray}
where $s\equiv |n-n'|$ and $\langle n|{\bf\Phi}|n\rangle =\pi/2$ for all $n$.
Now, in the classical limit $N=|\gamma|^{2}\rightarrow\infty$, 
$\langle n|\gamma\rangle\approx 0$ for $n\ll N$. Thus, the main contribution
to the sum in Eq.~(\ref{abba}) comes from the elements with $n\gg 1$.
Furthermore, since $\langle n+s|{\bf\Phi}|n\rangle\rightarrow 0$ for 
$s\rightarrow\infty$ the main contribution comes from elements with $n\gg s$.
Thus, we can approximate
\begin{equation}
\langle n+s|{\bf\Phi}|n\rangle = -\frac{1}{\pi}(1-(-1)^{s})
\left(\frac{1}{s^{2}}+\frac{1}{(2n+s)^{2}}\right)\approx
-\frac{1}{\pi}(1-(-1)^{s})\frac{1}{s^{2}}\;,
\end{equation}
and
\begin{equation}
\frac{|\gamma|^{2n}\gamma ^{s}}{\sqrt{n!(n+s)!}}\approx
\frac{|\gamma|^{2n}}{n!}\frac{\gamma ^{s}}{\sqrt{n^{s}}}\;,\;\;\;
\frac{|\gamma|^{2n}\gamma ^{-s}}{\sqrt{n!(n-s)!}}\approx
\frac{|\gamma|^{2n}}{n!}\frac{\sqrt{n^{s}}}{\gamma ^{s}}
\end{equation}
for $n\gg s$. Substituting these approximations in Eq.~(\ref{abba}) gives
\begin{eqnarray}
\langle\gamma|{\bf\Phi}|\gamma\rangle 
& \approx &
\frac{\pi}{2}-\frac{2}{\pi}\exp(-|\gamma|^{2})\sum_{n=0}^{\infty}\sum_{s=1,3,5...}
\frac{1}{s^{2}}\left[\frac{|\gamma|^{2n}}{n!}\frac{\gamma ^{s}}{n^{s/2}}+
\frac{|\gamma|^{2n}}{n!}\frac{n^{s/2}}{\gamma ^{s}}\right]\nonumber\\
& \approx & 
\frac{\pi}{2}-\frac{2}{\pi}\sum_{s=1,3,5...}\frac{1}{s^{2}}
\left[\frac{\gamma ^{s}}{N^{s/2}}+\frac{N^{s/2}}{\gamma ^{s}}\right],
\label{aba}
\end{eqnarray}
where $N\equiv\langle\gamma|{\bf N}|\gamma\rangle=|\gamma|^{2}$. We have used the 
fact that in the classical limit 
$\langle {\bf N}^{\pm s/2}\rangle\approx N^{\pm s/2}$.
Now, after substituting $\gamma ^{s}=N^{s/2}e^{i s\theta}$ in Eq.~(\ref{aba}),
we find that in the classical limit
\begin{equation}
\langle\gamma|{\bf\Phi}|\gamma\rangle \rightarrow \frac{\pi}{2}
-\frac{4}{\pi}\sum_{s=1,3,5...}\frac{\cos(\theta s)}{s^{2}} 
=|\theta|,
\label{thet}
\end{equation}   
since the r.h.s. is exactly the Fourier series of of the function
$f(\theta)=|\theta|$ ($-\pi<\theta<\pi$). This result proves useful for 
establishing that ${\bf\Phi}$ has the correct large-field correspondence 
limit. 

The agreement with the classical limit, given by Eq.~(\ref{thet}), can be shown
not just for the average of ${\bf \Phi}$ itself, but for the average of any 
analytical function of ${\bf \Phi}$. To show this we first calculate the 
average of $\sin {\bf \Phi}$. In the same way as Eq.~(\ref{thet}) was
obtained, it can be shown that
\begin{equation}
\langle\gamma|\sin {\bf\Phi}|\gamma\rangle= 
\frac{2}{\pi}+\exp(-|\gamma|^{2})\sum_{n=0}^{\infty}\sum_{s=1}^{\infty}
\left[\frac{|\gamma|^{2n}\gamma ^{s}}{\sqrt{n!(n+s)!}}+
\frac{|\gamma|^{2n}\gamma ^{-s}}{\sqrt{n!(n-s)!}}\right]
\langle n+s|\sin {\bf\Phi}|n\rangle,
\label{jay}
\end{equation}
since $\langle n|\sin {\bf\Phi}|n\rangle=2/\pi$. Now, it can be shown that 
for $n\gg s$
\begin{equation}
\langle n+s|\sin {\bf\Phi}|n\rangle\approx
-\frac{1+(-1)^{s}}{\pi}\frac{1}{s^{2}-1}.
\end{equation}
Substituting this in Eq.~(\ref{jay}) leads to  
\begin{equation}
\langle\gamma|\sin({\bf\Phi})|\gamma\rangle \rightarrow  \frac{2}{\pi}
-\frac{4}{\pi}\sum_{s=2,4,6...}\frac{\cos(\theta s)}{s^{2}-1}
=|\sin\theta|,
\end{equation}
in the limit $N\rightarrow\infty$. The r.h.s. is exactly 
the Fourier series of of the function $f(\theta)=|\sin\theta|$, so we
have the expected result.
Furthermore, Eq.~(\ref{1c}) and Eq.~(\ref{2s}) imply that for 
$N\rightarrow\infty$
\begin{eqnarray}
\langle\gamma|\cos({\bf\Phi})|\gamma\rangle & \rightarrow & \cos\theta\nonumber\\
\langle\gamma|\cos^{2}({\bf\Phi})|\gamma\rangle & \rightarrow & 
\cos^{2}\theta\nonumber\\
\langle\gamma|\sin^{2}({\bf\Phi})|\gamma\rangle & \rightarrow & 
\sin^{2}\theta
\label{clim}
\end{eqnarray}
since $\langle {\bf C}\rangle\rightarrow\cos\theta$ and 
$\langle {\bf S}\rangle\rightarrow\sin\theta$. Thus, all these results imply
that any power of $\sin{\bf \Phi}$ and $\cos{\bf \Phi}$ has the correct
classical limit, so that 
$\langle\gamma|f({\bf\Phi})|\gamma\rangle\rightarrow f(|\theta|)$ for
any analytic function $f(x)$.

\section{The quantum phase in a finite dimensional Hilbert space}

It is also interesting to discuss the quantum phase problem in a finite dimensional 
space~\cite{Pegg}. One can raise the question whether if instead of defining 
${\cal H}_{L}^{+}$, we could define a finite dimensional subspace of 
${\cal H}_{L}$ in which
the phase operator is well defined, and then take the limit of dimensionality
to infinity. Is this procedure equivalent to our earlier construction of 
${\cal H}_{L}^{+}$?   
  
According to our formalism, a finite dimensional Hilbert space can be obtained
from ${\cal H}_{L}$ by adding the requirement of periodicity in momentum 
space. That is, instead of working with ${\cal H}_{L}^{+}\subset{\cal H}_{L}$
we shall define a finite dimensional Hilbert space 
${\cal H}_{L}^{m}\subset {\cal H}_{L}$ as follows: for any 
$|\psi\rangle\in{\cal H}_{L}^{m}$ and for any $|p_{n}\rangle\in {\cal H}_{L}$, 
$\langle p_{n}|\psi\rangle =\langle p_{n+m+1}|\psi\rangle$ ($m$ is an integer). 
Thus, in the same way as the domain of $x$ has been restricted to (-L/2,L/2]
by the requirement of periodicity in the $x$-space, the domain of $p_{n}$, 
in ${\cal H}_{L}^{m}$, can be restricted to the range $n=0,1,2,...,m$.
That is, 
\begin{equation}
{\cal H}_{L}^{m}=span\{|p_{n}\rangle\}_{n=0}^{m}, 
\end{equation}
and therefore ${\cal H}_{L}^{m}$ is a finite Hilbert space of
dimension $m+1$.

Notice that the domain of $p_{n}$ could be chosen differently. For example, 
assuming $m$ is an even number, the domain of $p_{n}$ can be restricted
also to the range $n=-m/2,-m/2+1,...,m/2$. Using this range we can define
\begin{equation}
{\cal H}_{L}^{\pm m}=span\{|p_{n}\rangle\}_{n=-m/2}^{m/2}, 
\end{equation}
where the subscript ${}^{\pm}$ just indicates that $n$ can be also negative.
It is clear that ${\cal H}_{L}^{\pm m}\equiv {\cal H}_{L}^{m}$, i.e. both
Hilbert spaces are equivalent. However, as we shall see in the following,
the difference between the two Hilbert spaces appears in the limit process 
$m\rightarrow\infty$. 

The spectrum of ${\bf x}_{L}$ is no longer continuous since 
$\langle p_{n}|x\rangle =\langle p_{n+m+1}|x\rangle$. This requirement
implies that the spectrum of the position operator in 
${\cal H}_{L}^{m}\equiv {\cal H}_{L}^{\pm m}$ is given by
\begin{equation}
x_{l}=\frac{l}{m+1}L\;,\;\;\;\;l=-m/2,-m/2+1,...,m/2,
\end{equation}  
where we have used Eq.~(\ref{ccon}). Hence, the position ${\bf x}_{L}^{m}$ 
and momentum ${\bf p}_{L}^{m}$ operators
in ${\cal H}_{L}^{m}$ are given by
\begin{equation}
{\bf x}_{L}^{m}=\sum_{l=-m/2}^{m/2}x_{l}|x_{l}\rangle\langle x_{l}|\;\;\;
{\rm and}\;\;\;
{\bf p}_{L}^{m}=\sum_{n=0}^{m}p_{n}|p_{n}\rangle\langle p_{n}|\;\;\;
\left({\rm or}\;\;\;
{\bf p}_{L}^{m}=\sum_{n=-m/2}^{m/2}p_{n}|p_{n}\rangle\langle p_{n}|\right).
\end{equation}
The dimensionless angle operator is defined by 
${\bf\Theta}^{m}\equiv\frac{2\pi}{L}{\bf x}_{L}^{m}$ and the dimensionless
number (angular momentum) operator by 
${\bf N}^{m}=\frac{L}{2\pi\hbar}{\bf p}_{L}^{m}$.

We shall now examine the operators ${\bf\Theta}^{m}$ and ${\bf N}^{m}$ 
in the limit $m\rightarrow\infty$. This limit exists only in the case 
where $n=-m/2,-m/2+1,...,m/2$.
This can be proved by showing that the matrix elements of 
${\bf\Theta}^{m}$ in the momentum $|n\rangle$ basis are given by
\begin{equation}
\langle n'|{\bf\Theta}^{m}|n\rangle =\frac{(-1)^{n-n'}}{\frac{m+1}{2\pi}
\left[1-\exp\left(\frac{2\pi i(n'-n)}{m+1}\right)\right]}\;,
\end{equation}
for both ranges of $n$. However, in the limit $m\rightarrow\infty$ these
matrix elements coincide with the matrix elements of an angle observable
of plane rotator (see Eq.~(\ref{matp})). Thus, in the limit $m\rightarrow\infty$,
$\;{\bf\Theta}^{m}$ represents the canonical conjugate to the angular momentum (and 
not to the number operator). That is, $n$ must be unbounded from below.

Furthermore, if one is calculating some physical quantities in    
${\cal H}_{L}^{m}\equiv {\cal H}_{L}^{\pm m}$ (such as the variance of 
${\bf\Theta}^{m}$ in a momentum eigenstate) and then takes the limit 
$m\rightarrow\infty$, one will get the result obtained for a plane rotator,
$not$ for a linear harmonic oscillator. This explains why, for example,
finite dimensional quantum phase theories pass the Barnett and Pegg 
``acid-test''~\cite{Barnett}:
They just prove that an angular momentum (not number) eigenstate represents 
a state of indeterminate angle (not phase).

\section{Summary and conclusions}  

Only the absolute value of the phase observable of a single field mode  
is well defined. The definition of the phase turns out to be problematic
also from the classical point of view. It appears as a result of the fact
that the photon 
number is bounded from below. In the example of a plane rotator, the angle is well 
defined since the angular momentum can also have negative eigenvalues.

The phase is complementary to the excitation (or photon number) and thus 
the number eigenstates represent states of random phase. However, since 
only the absolute value of the quantum phase is well defined, number states
do not correspond to a uniform absolute phase distribution. 

The time operator which is the canonical
conjugate to the Hamiltonian of an harmonic oscillator 
is determined by: $|{\bf t}|=\omega ^{-1}{\bf\Phi}$.
Hence, our theory suggests that in order to define the complementary operator
for the Hamiltonian of a general system, one should seek the absolute value 
of the time observable.

\section*{Acknowledgments}
I would like to thank J.~Bekenstein 
for his guidance and support. This research was supported 
by grant No. 129/00-1 of the Israel Science Foundation,
and by a Clore Foundation fellowship.

\end{document}